\documentclass[twocolumn]{aastex63}

\received{----}
\revised{----}
\accepted{----}

\begin{document}

\title{What constraints on the neutron star maximum mass can one pose from GW170817 observations?}

\correspondingauthor{He Gao}
\email{gaohe@bnu.edu.cn}

\author{Shunke Ai}
\affiliation{Department of Physics and Astronomy, University of Nevada Las Vegas, Las Vegas, NV 89154, USA}
\affiliation{Department of Astronomy, Beijing Normal University, Beijing 100875, People's Republic of China}

\author{He Gao}
\affiliation{Department of Astronomy, Beijing Normal University, Beijing 100875, People's Republic of China}

\author{Bing Zhang}
\affiliation{Department of Physics and Astronomy, University of Nevada Las Vegas, Las Vegas, NV 89154, USA}

\begin{abstract}
The post-merger product of the first binary neutron star merger event detected in gravitational waves, GW170817, depends on neutron star equation of state (EoS) and is not well determined. We generally discuss the constraints one may pose on the maximum mass of a non-spinning neutron star, $M_{\rm TOV}$, based on the observations and some EoS-independent universal relations of rapidly-spinning neutron stars. If the merger product is a black hole after a brief hypermassive neutron star (HMNS) phase, we derive $M_{\rm TOV} < 2.09^{+0.11}_{-0.09}(^{+0.06}_{-0.04}) M_{\odot}$ at the 2$\sigma$ (1$\sigma$) level. The cases for a massive neutron star (MNS), either a supra-massive neutron star (SMNS) or even a stable neutron star (SNS), are also allowed by the data. We derive $2.09^{+0.11}_{-0.09}(^{+0.06}_{-0.04} M_{\odot}) \leq M_{\rm TOV}<  2.43^{+0.10}_{-0.08}(^{+0.06}_{-0.04}) M_{\odot}$ for the SMNS case and $M_{\rm TOV} \geq 2.43^{+0.10}_{-0.08}(^{+0.06}_{-0.04})M_{\odot}$ for the SNS case, at the $2\sigma$ ($1\sigma$) confidence level. In the MNS cases, we also discuss the constraints on the neutron star parameters (the dipolar magnetic field strength at the surface $B_p$ and the ellipticity $\epsilon$) that affect the spindown history, by considering different MNS survival times, e.g. 300 s, 1 d, and 155 d after the merger, as suggested by various observational arguments. We find that once an SMNS is formed, without violating the EM observational constraints, there always exist a set of ($B_p, \epsilon$) parameters that allow the SMNS to survive for 300s, 1 d, 155 d, or even longer. 
\end{abstract}


\section{introduction} 
On August 17, 2017, LIGO-Virgo Collaboration detected a GW signal from a binary neutron star (NS) merger event (GW170817). The total gravitational mass of the system at infinite binary separation is $2.74^{+0.04}_{-0.01}M_{\odot}$, with a mass ratio in the range of $(0.7-1)$ \citep{abbott17a}. Limited by the sensitivity of current GW detectors at high frequencies, the GW signals from the oscillations of the post-merger product are undetectable, leaving the merger product unidentified \citep{abbott17b,abbott19}. 

Theoretically, an NS-NS merger can in principle give a variety of post-merger products depending on the remnant (gravitational) mass ($M_{\rm rem}$, which depends on its spin)\footnote{Throughout the paper, without specification, masses refer to gravitational masses. The baryonic masses are denoted as $M_b$. For the same $M_b$, the gravitational mass can adopt a range of values, which depend on the spin state of the NS.} and the neutron star equation of state (EoS), which defines the maximum gravitational mass, $M_{\rm max}$, of an NS without collapsing into a black hole (BH) \citep{rezzolla10,bartos13,lasky14,ravi14,gao16, margalit17}. 
Given an EoS, the maximum mass of a non-rotational NS (denoted as $M_{\rm TOV}$) can be derived by solving the TOV equations \citep{oppenheimer39}. With rotation, the mass can be enhanced by a factor 
\begin{eqnarray}
\chi={(M-M_{\rm TOV}) \over M_{\rm TOV}}.
\label{eq:chi}
\end{eqnarray} 
In the following, $(1+\chi)$ is defined by the enhancement factor due to uniform rotation. Differential rotation can also enhance the mass, and we denote the enhancement factor as $(1+\chi_d)$. Typically one has $\chi < \chi_d$. 
The degree of enhancement has been extensively studied in the literature \citep{cook94,lasota96,breu16,studzinska16,gondek17,bauswein17,bozzola18,weih18}. 
One can define 
\begin{eqnarray}
    M_{\rm max} & \equiv & (1+\chi_{\rm max}) M_{\rm TOV}, \label{eq:Mmax} \\
    M_{\rm max,d} & \equiv & (1+\chi_{\rm d,max}) M_{\rm TOV},
\end{eqnarray}
where $\chi_{\rm max} \sim 0.2$ \citep{cook94,lasota96,breu16} and $\chi_{\rm d,max} \sim (0.3-0.6)$
\footnote{The quoted $\chi_{\rm d,max}$ range is roughly derived from the numerical result that $M_{\rm tot}>kM_{\rm TOV}$ (where $M_{\rm tot}=M_1+M_2$ and $k=1.3-1.6$) is required to directly form a BH as the merger remnant \citep{shibata05,shibata06,hotokezaka11,bauswein13,margalit17}. Since $M_{\rm TOV}<M_{\rm tot}$, the true $\chi_{\rm d,max}$ should be somewhat smaller than 0.3-0.6.}
are the maximum enhancement factor for uniform and differential rotations, respectively. Let us define $M_{\rm rem}^0$, $M_{\rm rem}^k$, and $M_{\rm rem}^\infty$ as the gravitational masses of the remnant right after the merger, at the beginning of uniform rotation (which is assumed to carry a Keplerian rotation), and with no rotation ($P=\infty$), respectively. The fate of the merger product can be then determined as follows: If $M_{\rm rem}^0 > M_{\rm max,d}$, the merger remnant would directly collapse to a BH. Otherwise the remnant would go through a hypermassive NS (HMNS) phase,  during which the merger remnant loses angular momentum as well as mass. If $M_{\rm max} < M_{\rm rem}^k < M_{\rm rem}^0 \leq M_{\rm max,d}$, the merger remnant would collapse into a BH after the HMNS phase. If, however, $M_{\rm rem}^k \leq M_{\rm max}$, a uniformly rotating NS would be formed after the differential rotation is damped. Whether it is an SMNS or an SNS depends on the comparison between $M_{\rm TOV}$ and $M_{\rm rem}^{\infty}$. If $M_{\rm rem}^\infty > M_{\rm TOV}$, the merger remnant is a supramassive NS (SMNS), which would eventually collapse into a BH. If $M_{\rm rem}^\infty \leq M_{\rm TOV}$, the remnant would never collapse, which is a stable NS (SNS). Either an SMNS or an SNS can be called as a massive NS (MNS). 

For GW170817, although GW data cannot determine the nature of the merger product, it has been suggested that the EM counterpart observations may provide some clues. Unfortunately, owing to the messy physics involved in producing EM counterparts, all the claims on the constraints on $M_{\rm TOV}$ rely on some assumptions, so that no consensus can be reached. All investigators agree on that the remnant cannot be a promptly formed black hole, since the observed kilonova is too bright to be explained by the dynamical ejecta only. The disagreement comes to the lifetime of the NS produced during the merger. Many authors assumed that in order to produce the short gamma-ray burst (GRB 170817A) \citep{goldstein17,zhangbb18} following GW170817, a BH engine is needed \citep[e.g.][]{margalit17,rezzolla18,ruiz18}. Within this picture, the remnant is an HMNS, which must have collapsed before the GRB trigger time, which is 1.7 s after the merger \citep{abbott17c,goldstein17}. In particular, a good fraction of the observed 1.7 s delay has to be attributed to the HMNS phase, during which significant mass ejection is warranted to account for the observed bright kilonova emission  \citep{siegel17,gill19}.
If this is the case, the multi-messenger observations of GW170817 can be used to provide an upper bound on $M_{\rm TOV}$ \citep{margalit17,rezzolla18,ruiz18,shibata19}, e.g. $M_{\rm TOV}\lesssim2.16~M_{\odot}$ by \cite{margalit17}. 

On the other hand, in order to explain the extended engine activities (flares, extended emission, and internal X-ray plateaus) of short GRBs, it has been long proposed that at least some NS-NS merger systems can produce both a short GRB and an MNS \citep{dai06,gao06,metzger08,dessart09,lee09,zhang10,fernandez13}. Indeed, when interpreting the rapid decay at the end of internal X-ray plateau observed in a good fraction of short GRBs \citep{rowlinson10,rowlinson13,lu14,lu15}, $M_{\rm TOV}$ has to be (much) greater than $2.16~M_{\odot}$\citep{gao16,li16}. So there is a direct conflict between the upper limit on $M_{\rm TOV}$ derived from GW170817 (assuming that a BH is formed before 1.7 s after the merger) and the short GRB X-ray plateau data. Indeed, since GRB 170817A did not trigger Swift, there was no early X-ray afterglow data to check whether there was an early X-ray plateau phase similar to other short GRBs. The fact that the delay time (1.7 s) is comparable to the GRB duration itself ($\sim 2$ s) also suggests that the jet launching waiting time $\Delta t_{\rm jet}$ as well as the shock breakout time $\Delta t_{\rm bo}$ may be small \citep{zhang19}. If so, the launch of a jet may not demand a BH, and the bright kilonova emission may benefit from energy injection of a long-lived remnant \citep{yu18,li18}. Indeed, the ``blue'' component of the kilonova peaks at $\sim 1$ d  with a peak luminosity  $\sim 10^{42}{\rm erg~s^{-1}}$, which requires a $>0.02M_{\odot}$ mass of lanthanide-free ejecta ($Y_e \gtrsim 0.25$) with $v_{\rm ej,blue}\approx 0.2-0.3c$ \citep{kasen17,cowperthwaite17,chornock17,kilpatrick17,villar17,tanvir17,gao17,shappee17}. In order to fit both the peak time and peak luminosity, a small opacity $\kappa \sim (0.3-0.5) \ {\rm cm^2 \ g^{-1}}$ is required, which is in conflict with the value $\kappa \sim 1 \ {\rm cm^2 \ g^{-1}}$ derived from the most detailed calculations for Fe group elements \citep{tanaka19}. Energy injection from a long-lived MNS that survives for at least 1 day can help to ease the conflict and interpret the blue component \citep{li18}. Finally, \cite{piro19} claimed a low-significance temporal feature at 155 days in the X-ray afterglow of GW170817 which carries properties of GRB X-ray flares. If such a feature is not due to a statistical fluctuation, one would demand an active central engine at such a late epoch. The putative MNS should at least survive for 155 days after the merger.

In this work, instead of claiming the identity of  the merger remnant of GW170817, we leave it as an open question and generally discuss what constraints on $M_{\rm TOV}$ one can place for different possible merger remnants. We estimate the remnant mass of GW170817 based on the gravitational wave data and NS EoSs (either for individual EoSs or using some EoS-independent universal relations). We identify the separation lines among three types of products (HMNS/BH, SMNS, and SNS) in Section 2. In Section, 3, we discuss the cases of SMNS and SNS and place constraints on the NS parameters (surface magnetic field at the pole, $B_p$, and ellipticity, $\epsilon$) assuming that the remnant can survive for 300 s (typical ending time of internal X-ray plateaus), 1 d (peak time of the blue kilonova component), and 155 d (time of the putative X-ray flare), respectively.

\section{Constraints on $M_{\rm TOV}$ for different types of merger product}

\subsection{General approach}\label{sec:general-approach}

Our goal is to address the following question: given the information provided by the gravitational wave data from GW170817, i.e. the total gravitational mass at infinite binary separation $M_{\rm tot}= M_1 + M_2 = 2.74^{+0.04}_{-0.01}M_{\odot}$ and mass ratio $q = m_1/m_2 = (0.7-1)$ under the low dimensionless NS spin prior \citep{abbott17a}, what can one say about the maximum mass of the non-spinning NS $M_{\rm TOV}$? Our general approach is as follows:
\begin{itemize}
    \item Assume the values of $M_1$ and $M_2$. In our case, we assume that $M_1 = M_2 = M_{\rm tot} / 2$, noticing that the total baryonic mass is highly insensitive to the binary mass ratio;
    \item Convert the gravitational masses to baryonic masses $M_{1,b}$ and $M_{2,b}$, either based on the public code \texttt{RNS} code (for individual EoSs) \citep{stergioulas95} or some EoS-independent universal relations \citep[e.g.][]{gao19};
    \item {Conserve the total baryonic mass of the binary system throughout the merger. In the post-merger phase, the baryonic mass in the remnant could be derived by subtracting the baryonic mass of various ejecta components from the total baryonic mass}, i.e. $M_{\rm rem,b}=(M_{1,b} + M_{2,b}) - M_{\rm ejc}$. Based on the EM counterpart observations, the total ejected mass is estimated as $M_{\rm ejc} \sim (0.06\pm0.01) M_\odot$ \cite[][and reference therein]{metzger17};
    \item Convert $M_{\rm rem,b}$ to the gravitational mass of the central object $M_{\rm rem}$, which depends on its spin state \citep{gao19}. In particular, we care mostly about $M_{\rm rem}^k$ and $M_{\rm rem}^\infty$.
    \item Compare $M_{\rm rem}^k$ against $M_{\rm max}$
    or 
    \begin{equation}
      M_{\rm TOV}^k = (1+\chi_{\rm TOV}^k) M_{\rm TOV}  
      \label{eq:MTOVk}
    \end{equation}
    to determine whether the final merger product is an HMNS/BH, SMNS, or SNS. Here $M_{\rm TOV}^k$ is the gravitational mass at Keplerian rotation for an NS whose non-spin gravitational mass is $M_{\rm TOV}$.
\end{itemize}

In the following, we discuss the results for individual EoSs (\S2.2) and for general cases using universal relations (\S2.3).

\subsection{Individual EoSs}
We adopt 10 realistic (tabulated) EoSs (as listed in Table 1): SLy \citep{douchin01}, WFF1 \citep{wiringa88}, WFF2 \citep{wiringa88}, AP3 \citep{akmal97}, AP4 \citep{akmal97}, BSK21 \citep{goriely10}, DD2 \citep{typel10}, MPA1 \citep{muther87}, MS1 \citep{muller96}, MS1b \citep{muller96} with $M_{\rm TOV}$ ranging from $2.05M_{\odot}$ to $2.78M_{\odot}$. For each EoS, we use \texttt{RNS} to calculate $M_{\rm TOV}$. We also calculate the allowed minimum Keplerian period (marked as $P_{\rm k,min}$), which is related to the Keplerian period of the NS at $M_{\rm max}$. The results are collected in Table 1.

For each EoS, we derive the $M_{\rm rem,b}$ following the approach described in Section \ref{sec:general-approach}. Since the $M_b-M$ relation is somewhat different for different EoSs, the derived $M_{\rm rem,b}$ is different even for the same event GW170817. With the derived $M_{\rm rem,b}$ for each EoS, we apply the \texttt{RNS} code to see whether there is a uniformly rotating NS solution. If not, the merger product would be a BH (likely preceded by a brief HMNS phase). If a solution is available, we further test whether there is a solution in the non-spinning case. The remnant would be an SMNS or SNS if the answer is ``no'' or ``yes'', respectively. As shown in Table 1, among the 10 EoSs studied, one (SLy) with $M_{\rm TOV} = 2.05 M_{\odot}$ forms an HMNS/BH, three (MPA1, Ms1, Ms1b) with minimum $M_{\rm TOV} = 2.48 M_{\odot}$ (MPA1) form an SNS, and the other six (with $M_{\rm TOV}$ between $2.14M_{\odot}$ (WFF1) and $2.42M_{\odot}$ (DD2)) form an SMNS. According to this small sample investigation, the $M_{\rm TOV}$ separation line for HMNS/BH and SMNS may be between $2.05 M_{\odot}$ and $2.14M_{\odot}$, and that between SMNS and SNS may be between $2.42M_{\odot}$ and $2.48 M_{\odot}$.

\begin{table*}                                                                      
\begin{center}{\scriptsize                                                                                       
\caption{The 10 EoSs investigated in this paper.}                                                                             
\begin{tabular}{cccccccccc}                                                                                          
\hline                                                                                                                   
\hline                                                                                                                                                                                                               
    & $M_{\rm TOV}$            &$P_{\rm k,min}$   &$M_{\rm b,tot}$               &$M_{\rm b,rem}$              &$M_{\rm rem}^k$          &$P_k$      &$1+\chi_{\rm TOV}^{k}$      &$1+\chi_{\rm max}$     &Product type      \\
    &$\left(M_{\odot}\right)$  &${\rm ms}$        &$\left(M_{\odot}\right)$      &$\left(M_{\odot}\right)$     &$\left(M_{\odot}\right)$ &${\rm ms}$ &                        &                 &                  \\
\hline                                                                                                                                                                                                               
SLy &2.05                      &$0.55$            &$3.01^{+0.05}_{-0.01}$                          &$2.95^{+0.06}_{-0.02}$                           &$--$                     &$--$       &1.039                   &1.184            &BH                 \\
\hline                                                                                                                                                                                                               
WFF1&2.14                      &$0.47$          &$3.07^{+0.05}_{-0.01}$                               &$3.01^{+0.06}_{-0.02}$                        &$2.51^{+0.03}_{-0.01}$                     &$0.52$     &1.051                   &1.201            &SMNS               \\
\hline                                                                                                                                                                                                               
WFF2&2.20                      &$0.50$          &$3.04^{+0.05}_{-0.01}$                               &$2.98^{+0.06}_{-0.02}$                         &$2.51^{+0.04}_{-0.01}$                    &$0.58$     &1.048                   &1.192            &SMNS               \\
\hline                                                                                                                                                                                                               
Ap4 &2.22                      &$0.51$            &$3.03^{+0.05}_{-0.01}$                            &$2.97^{+0.06}_{-0.02}$                        &$2.52^{+0.03}_{-0.01}$                     &$0.60$     &1.047                   &1.194            &SMNS               \\
\hline                                                                                                                                                                                                               
BSk21&2.28                     &$0.60$          &$2.99^{+0.05}_{-0.01}$                               &$2.93^{+0.06}_{-0.02}$                         &$2.54^{+0.03}_{-0.02}$                     &$0.74$     &1.044                   &1.205            &SMNS               \\
\hline                                                                                                                                                                                                               
AP3 &2.39                      &$0.55$            &$3.01^{+0.05}_{-0.01}$                             &$2.95^{+0.06}_{-0.02}$                         &$2.54^{+0.03}_{-0.02}$                     &$0.70$     &1.049                   &1.202            &SMNS               \\
\hline                                                                                                                                                                                                               
DD2 &2.42                      &$0.65$            &$2.99^{+0.05}_{-0.01}$                             &$2.93^{+0.06}_{-0.02}$                        &$2.55^{+0.04}_{-0.01}$                     &$0.82$     &1.042                   &1.208            &SMNS               \\
\hline                                                                                                                                                                                                               
MPA1&2.48                      &$0.59$           &$3.00^{+0.05}_{-0.01}$                              &$2.94^{+0.06}_{-0.02}$                        &$2.54^{+0.04}_{-0.01}$                     &$0.76$     &1.048                   &1.208            &SNS               \\
\hline                                                                                                                                                                                                               
Ms1 &2.77                      &$0.72$            &$2.95^{+0.05}_{-0.01}$                             &$2.89^{+0.06}_{-0.02}$                         &$2.56^{+0.04}_{-0.01}$                    &$1.00$     &1.043                   &1.207            &SNS                \\
\hline                                                                                                                                                                                                               
Ms1b&2.78                      &$0.71$           &$2.96^{+0.05}_{-0.01}$                             &$2.90^{+0.06}_{-0.02}$                         &$2.56^{+0.04}_{-0.01}$                     &$0.99$     &1.042                   &1.212            &SNS                \\
\hline                                                                                                                                                                                                               
\hline                                                                          
 \end{tabular}                                                                      
 }                                                                                                                                                              
\end{center}                                                                        
\end{table*} 

\subsection{Universal Approach}

Since there are many more EoSs discussed in the literature \citep[e.g.][]{lattimer12}, it is impossible to make a self-consistent check for all the proposed EoSs. Some general constraints on $M_{\rm TOV}$ (with large uncertainties) may be obtained by applying some EoS-independent empirical relations for GW170817. 

\begin{figure}[ht!]
\resizebox{90mm}{!}{\includegraphics[]{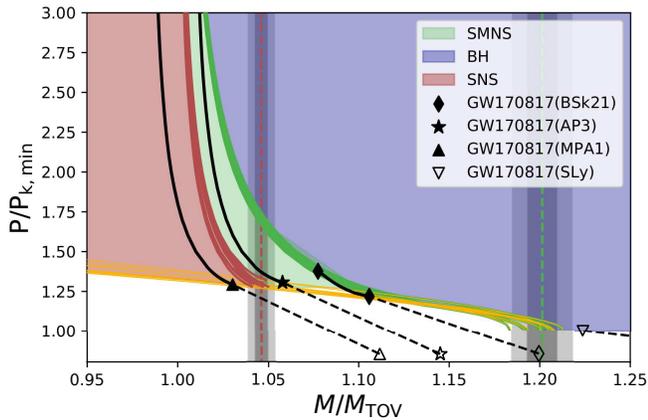}} 
\caption{The allowed parameter space of different types of merger products. The orange bunch of lines denote the mass-dependent normalized Keplerian period ${\cal P}_k$; the red bunch of lines denote of constant $M_{\rm TOV,b}$ lines; the green bunch of lines denote the boundary line for the SMNS to collapse into a BH. Each bunch includes 10 lines corresponding to 10 different EoSs. The black lines stand for the evolving trajectory of the merger product of GW170817. The markers show the points where the evolution phase changes. The dashed lines and hollow markers are schematic since the period of differential rotating NS is undefined. The vertical dark regions denote the separation lines for three regions at $P=P_k$.} 
\label{fig:type}
\end{figure}

Generally, the type of the merger product is best determined by comparing $M_{\rm rem}^k$ with $M_{\rm max}$ (Eq.(\ref{eq:Mmax})) and $M_{\rm TOV}^k$ (Eq.(\ref{eq:MTOVk})), both are highly dependent on the EoS. Fortunately, when the gravitational mass of an NS is normalized to $M_{\rm TOV}$ and when the rotation period $P$ is normalized to $P_{\rm k,min}$, the evolution of the separation boundaries among HMNS/BH, SMNS, and SNS in the ${\cal M} - {\cal P}$ plane (where ${\cal M} \equiv M / M_{\rm TOV} \equiv (1+\chi)$ and ${\cal P} \equiv P / P_{\rm k,min}$) is highly EoS-insensitive. This is shown in Figure \ref{fig:type}. The orange bunch of lines denote the Keplerian lines, which denote the normalized Keplerian period ${\cal P}_k$ as a function of the gravitational mass of the NS at that period. For the EoSs we investigate, one can get the best-fit line as 
\begin{eqnarray}
{\cal P}_k&=&(-2.697\pm0.355)\times({M \over M_{\rm TOV}})^2 \nonumber\\
&+&(4.355\pm0.764)\times({M \over M_{\rm TOV}})\nonumber\\
&-&(0.303\pm0.409),~~P>P_{\rm k,min}.
\end{eqnarray}
 One can see that $P_k$ becomes progressively longer when $M < M_{\rm max}$. This line is the starting point for the evolution of any NS after the differentiation rotation is damped. 

Let us now consider a spinning NS. Its baryonic mass never changes with the spin period while the gravitational mass would decrease as it spins down. These constant $M_b$ curves are examplified as the two black lines (for two particular $M_b$ values) and the red bunch of lines, which show constant $M_{\rm TOV,b}$ lines for different EoSs. The best-fit line for the 10 EoSs leads to
\begin{eqnarray}
\log_{10}\chi_{\rm TOV}&=&(1.804\pm0.268)\times({\rm log_{10}}{\cal P})^2 \nonumber\\
&+&(-3.661\pm0.190)\times{\rm log_{10}}{\cal P} \nonumber\\
&+&{\rm log_{10}(0.101\pm0.007)}, ~P>P_{\rm k,TOV}.
\end{eqnarray}
The intersection of this line and the best-fit orange line gives $P_{\rm k,TOV}$, which represents the Kepler period when $M_b=M_{\rm b,TOV}$, and the enhancement factor is $(1+\chi_{\rm TOV}^k)$.

A uniformly rotating NS with $M_b>M_{\rm TOV,b}$ would eventually collapse into a BH when $M >(1+\chi_{\rm col})M_{\rm TOV}$, where $\chi_{\rm col}$ is defined as the maximally allowed enhancement gain at a particular $P$. The values of $(1+\chi_{\rm col})$ with different $P$ values can serve as the separation line between SMNS and HMNS/BH regimes. This corresponds to the green bunch of lines in Figure \ref{fig:type}, which corresponds to the best fit as 
\begin{eqnarray}
\log_{10}\chi_{\rm col}&=&(-2.740\pm0.045)\times{\rm log_{10}} {\cal P} \nonumber\\
&+&{\rm log_{10}(0.201 \pm 0.005)}
\label{eq:chi_ucol}
\end{eqnarray}
When $P \rightarrow P_{\rm k,min}$, one has $\chi_{\rm col}^k \rightarrow \chi_{\rm max}$.

The region below the orange bunch of lines in Figure \ref{fig:type} (the white region) is not well defined, since $P$ cannot be defined for an differentially rotating object. The \texttt{RNS} code we employ can only be used in uniformly rotating case. We therefore indicate the evolution trajectories in the white region using dashed lines.

We plot several evolutionary trajectories of the GW170817 remnant within the framework or several EoSs: BSk21 (diamond), AP3 (star), MPA1 (upward triangle), and SLy (downward triangle). The symbols are solid or open in the uniformly or differentially rotating regimes, respectively. 

The mass at the starting point of rigid rotation, i.e. $M_{\rm rem}^k$, is crucial to determine the remnant type through its comparison with $M_{\rm max}$ (Eq.(\ref{eq:Mmax})) and $M_{\rm TOV}^k$ (Eq.(\ref{eq:MTOVk})). 
The values of $\chi_{\rm TOV}^k$ and $\chi_{\rm max}$ of each EoS can be calculated utilizing the \texttt{RNS} code, and may be also generally estimated as $\chi_{\rm TOV}=0.046 \pm 0.008 (\pm 0.004)$ and
$\chi_{\rm max}=0.201\pm 0.017 (\pm 0.008)$ with 2$\sigma$ (1$\sigma$) errors, respectively.

\begin{figure}[ht!]
\resizebox{90mm}{!}{\includegraphics[]{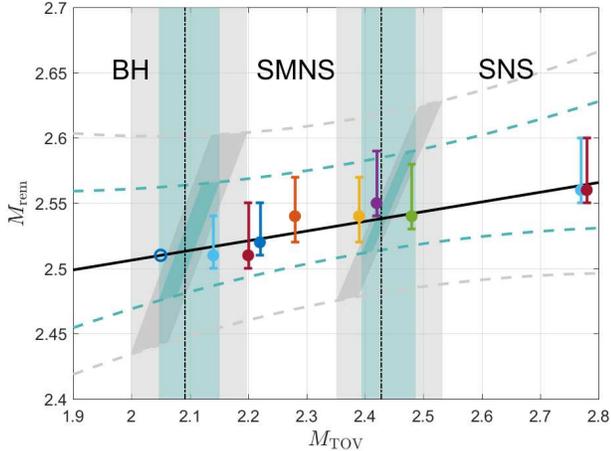}} 
\caption{Constraints on the range of $M_{\rm TOV}$ for three different merger products in the case of GW170817. The colored data points represent the values of $M_{\rm TOV}$ and the estimated $M_{\rm rem}$ at the Keplerian period for different EoSs. The solid line is the best fitting relation of $M_{\rm TOV}$ and $M_{\rm rem}$, whereas the blue and grey dashed lines showing the $1\sigma$ and $2\sigma$ error range, respectively.  The slanted deep grey shadows are the allow regions when $\chi=\chi_{\rm TOV}^k$ and $\chi=\chi_{\rm max}$, respectively. The dot-dashed vertical lines are the central $M_{\rm TOV}$ values of the separation lines of different merger products, which are surrounded by the $1\sigma$ (light blue shadow) and $2\sigma$ (light grey shadow) regions. The hollow circle is the predictive value for the EoS SLy, which forms a BH rather than a MNS.} 
\label{fig:M}
\end{figure}

As shown in Table 1, the value of $M_{\rm rem}^k$ only weakly depends on EoSs and is slightly correlated with $M_{\rm TOV}$. We show the relationship between $M_{\rm rem}^k$ and $M_{\rm TOV}$ in Figure \ref{fig:M}, which reads \begin{eqnarray}
M_{\rm rem}^{k}=(2.354 \pm 0.074)+(0.076\pm 0.032)M_{\rm TOV},
\label{eq:Mg}
\end{eqnarray}
with 2$\sigma$ error. Combining Equations \ref{eq:chi} and \ref{eq:Mg}, one can derive the critical values to separate the HMNS/BH vs. SMNS and SMNS vs. SNS, as also shown in Figure \ref{fig:M}. 

The following results can be obtained for GW170817: If the merger remnant is a HMNS/BH, $M_{\rm TOV}$ should be smaller than $2.09^{+0.11}_{-0.09}(^{+0.06}_{-0.04})M_{\odot}$ at the 2$\sigma$ (1$\sigma$) level\footnote{After mass shedding from the initial torus due to viscous and neutrino cooling processes, a quasi-stationary torus is supposed to exist surrounding the central core \citep{siegel17,hanauske17,fujibayashi18}. The mass remaining in the torus is in Keplerian orbits, thus would not add to the gravitational mass in the core. Considering this effect, the values of the critical $M_{\rm TOV}$ to separate the HMNS/BH and SMNS cases would be smaller. However, since a significant fraction of mass in the torus has fallen back or ejected within $\sim 0.3s$ \citep{siegel17,fujibayashi18} (comparable to the lifetime of HMNS \citep{metzger18}), the influence of the quasi-stationary torus to the critical $M_{\rm TOV}$ is small.
}; if the merger remnant is an SMNS, $M_{\rm TOV}$ should be in the range from 
$2.09^{+0.11}_{-0.09}(^{+0.06}_{-0.04})M_{\odot}$ to 
$2.43^{+0.10}_{-0.08}(^{+0.06}_{-0.04})M_{\odot}$ at the 2$\sigma$ (1$\sigma$) level; if the merger remnant is a SNS, $M_{\rm TOV}$ should be greater than $2.43^{+0.10}_{-0.08}(^{+0.06}_{-0.04}M_{\odot})$ at the 2$\sigma$ (1$\sigma$) level. These results are generally consistent with previous results assuming an HMNS/BH remnant in GW170817 \citep[e.g.][]{margalit17,ruiz18,rezzolla18,shibata19}, even though some details differ. For example, \cite{ruiz18} and \cite{rezzolla18} both adopted the measured $M_{\rm tot} \sim 2.74 M_\odot$ deducting the mass loss to estimate $M_{\rm rem}$. This over-estimated $M_{\rm rem}$ by $\sim 0.2 M_\odot$, which would over-estimate the upper limit of $M_{\rm TOV}$. \cite{shibata19} performed the most detailed analysis numerically and derived a more conservative upper limt $\sim 2.3 M_\odot$ for the HMNS/BH case, but for the majority of the EOSs studied, the range of $M_{\rm TOV}$ that form an HMNS/BH product is still consistent with our estimate. Our derived separation line between SMNS and SNS products is also consistent with theirs.

\section{Constraints on the NS Properties in the MNS cases}

In the case of SMNS, it would be interesting to investigate under what conditions the SMNS can survive for a particular duration of time, e.g. $\sim$ 300 s for the typical duration of the X-ray internal plateau, $\sim$ 1 d to power the blue component of the kilonova, and $\sim$ 155 d to power the putative X-ray flare. This depends on the NS EoS and the spindown history of the putative MNS remnant.

At a certain spin period $P$, $\chi_{\rm TOV}<\chi<\chi_{\rm col}$, the remnant would be an SMNS. For a particular object, $\chi$  decreases as the NS loses its rotation energy through magnetic dipole radiation and GW radiation \citep{shapiro83,zhang01}, i.e.
\begin{eqnarray}
\dot{\Omega}=-\frac{32GI\epsilon^2\Omega^5}{5c^5}-\frac{B_p^2R^6\Omega^3}{6Ic^3},
\label{dotE}
\end{eqnarray} 
where $\Omega=2\pi /P$ is the angular frequency and $\dot{\Omega}$ is its derivative, $B_p$ stands for the surface dipole magnetic field strength at the pole, and $\epsilon$ presents the ellipticity of the NS. As the NS spins down, it would collapse into a BH when $\chi>\chi_{\rm col}$.

For a rigidly rotating NS,  the maximum enhancement factor $\chi_{\rm col}$ at any given $P$ can be estimated with the EoS-independent relation Equation (\ref{eq:chi_ucol}), which could be up to $\chi_{\rm max} \sim 20\%$ when the NS spin period equals to the allowed minimum Keplerian period. Given the initial spin period (Keplerian) and a particular desired lifetime of the MNS, it is possible to follow the spindown evolution and constrain $B_p$ and $\epsilon$. 

\begin{figure*}[ht!]
\begin{center}
\begin{tabular}{l}
\resizebox{115mm}{!}{\includegraphics[]{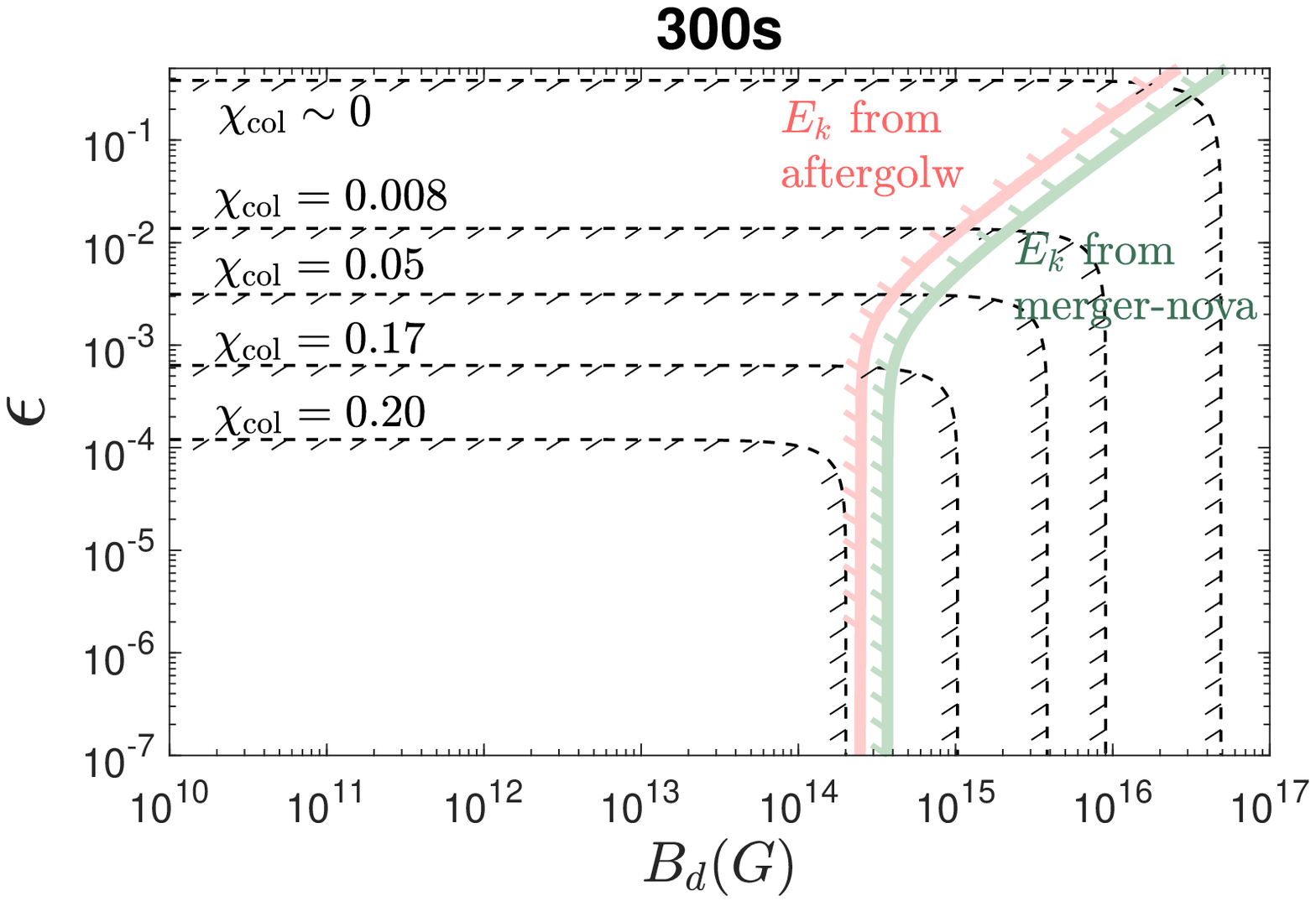}} \\ 
\resizebox{115mm}{!}{\includegraphics[]{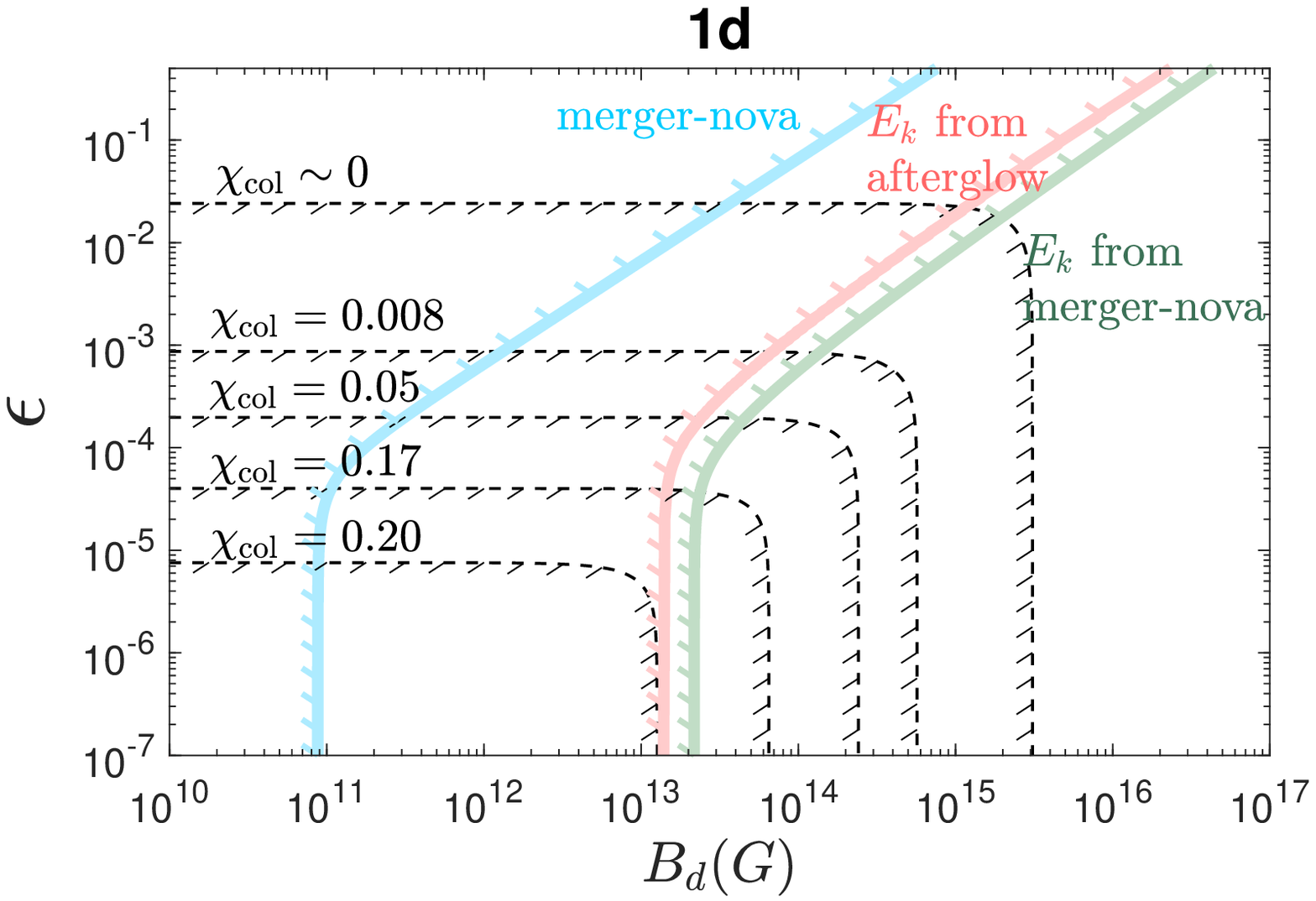}} \\
\resizebox{115mm}{!}{\includegraphics[]{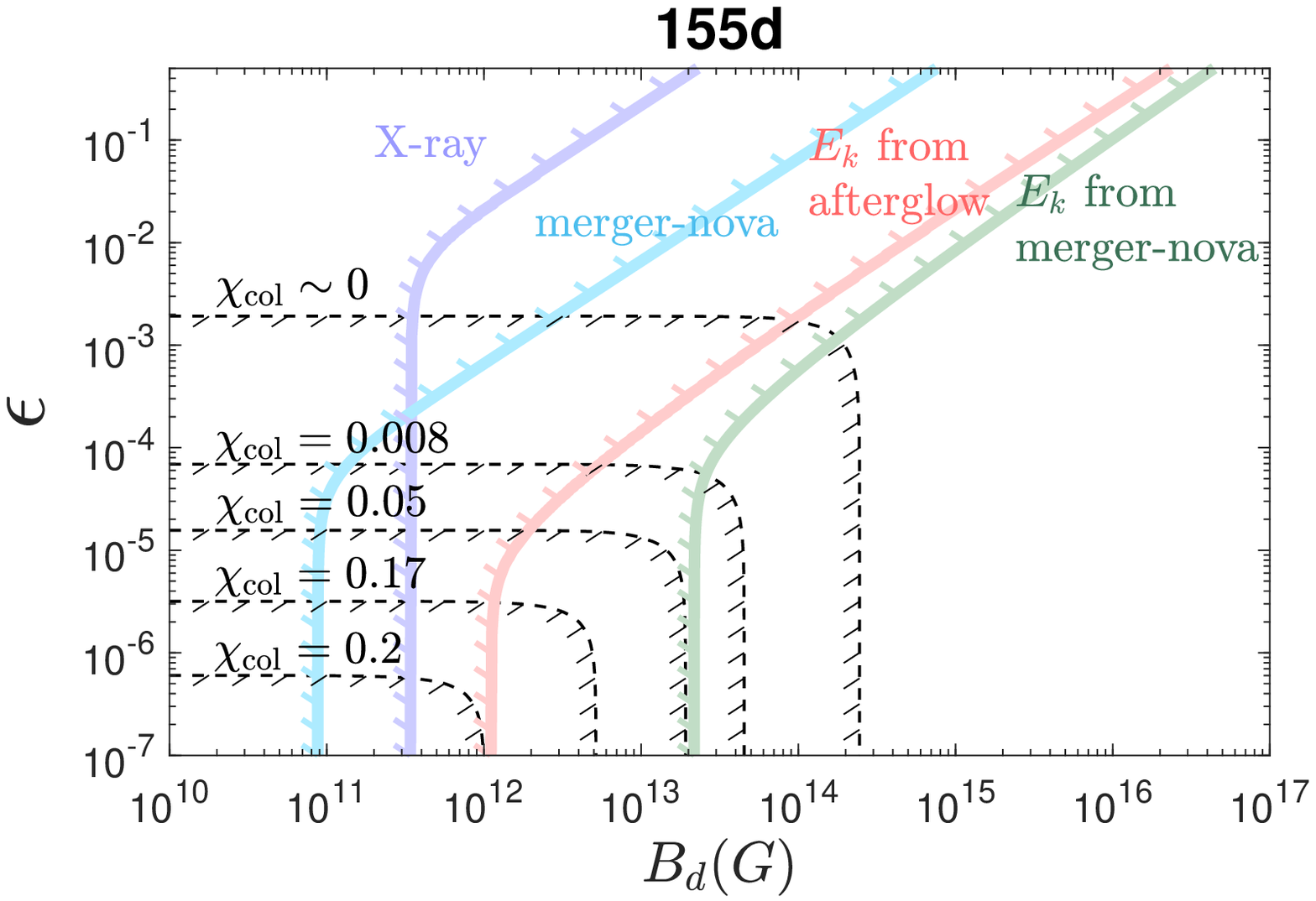}} 
\end{tabular}
\caption{Constraints on the allowed parameter space in the $B_p$-$\epsilon$ plane for three survival times: 300 s (upper), 1 d (middle), and 155 d (lower).
The dashed lines are the boundaries for different putative EoSs (which correspond to different $\chi$) within which the SMNS can survive for a certain timescale. The red and green lines show the constraints from the kinetic energy of ejecta, which are deduced from observations of afterglow and mergernova, respectively. The blue and purple lines stand for the constraints from the luminosity of the mergernova and the late-time X-ray signal. }
\label{fig:universal}
\end{center}
\end{figure*}

To be conservative, we assume the initial period $P_i=P_{\rm k,min}$ for GW170817 and use the moment of inertia and radius of a non-rotating NS to constrain $B_p$ and $\epsilon$ parameters\footnote{The initial spin period of the merger product, $P_k$ should not be smaller than $P_{\rm k,min}$, which means that smaller $\epsilon$ and $B_p$ values than constrained are required for the SMNS to spin down to a certain period. Our derived upper limits on $B_p$ and $\epsilon$ are therefore safe upper limits. Similarly, if one take larger $I$ and $R$ values for spinning NSs, one also requires smaller $B_p$ and $\epsilon$ values than derived to reach the same spindown effect.}. Given a value of $\chi_{\rm col}$, we plot the boundary lines in the $B_p-\epsilon$ plane for the region which allows an SMNS to survive for a certain lifetime, e.g. 300 s, 1 day and 155 days (see Figure \ref{fig:universal}). During a particular time span, part of the spin-down power of the SMNS would be released in the EM channel, so that the EM counterpart observations of GW170817 could be used to make constraints on the spin-down power of the remnant SMNS, and hence,  on $B_p$ and $\epsilon$. 

If the SMNS can survive for 300s, we consider two constraints from the EM counterpart observations: the EM channel spin-down power integrated within 300s should be less than the kinetic energy of the merger ejecta [$\beta<0.3$, inferred from the spectrum observation of the optical counterpart \citep{kasen17,cowperthwaite17,chornock17,kilpatrick17,shappee17}] and less than the kinetic energy of the GRB jet [$E_k<10^{51}{\rm erg~s^{-1}}$, inferred from the radio afterglow emission \citep{nakar18,dobie18}]. If the SMNS can survive for 1 d, in addition to the constraints from the kinetic energy, since the merger ejecta already became optically thin at that time, the luminosity of the magnetic dipole spin-down should be smaller than the peak luminosity of the optical counterpart. If the SMNS can survive for 155d, besides the constraints from the kinetic energy and optical peak luminosity, the late time X-ray observations could also serve as the upper limit of the luminosity of the magnetic dipole spin-down. These constraints on $B_p$ and $\epsilon$ from EM counterpart observations of GW170817 are  shown in Figure \ref{fig:universal}.  We can see that the EM observations tend to constrain $B_p$ and $\epsilon$ to small values. Nonetheless, if an SMNS is formed, there always exits a suitable $(B_p, \epsilon)$ parameter space to allow the SMNS to survive for 300 s, 1 d, 155 d, or even longer,
without violating the observational constraints.
These constraints are clearly displayed in Figure \ref{fig:universal}.

\begin{figure*}[ht!]
\begin{center}
\begin{tabular}{l}
\resizebox{115mm}{!}{\includegraphics[]{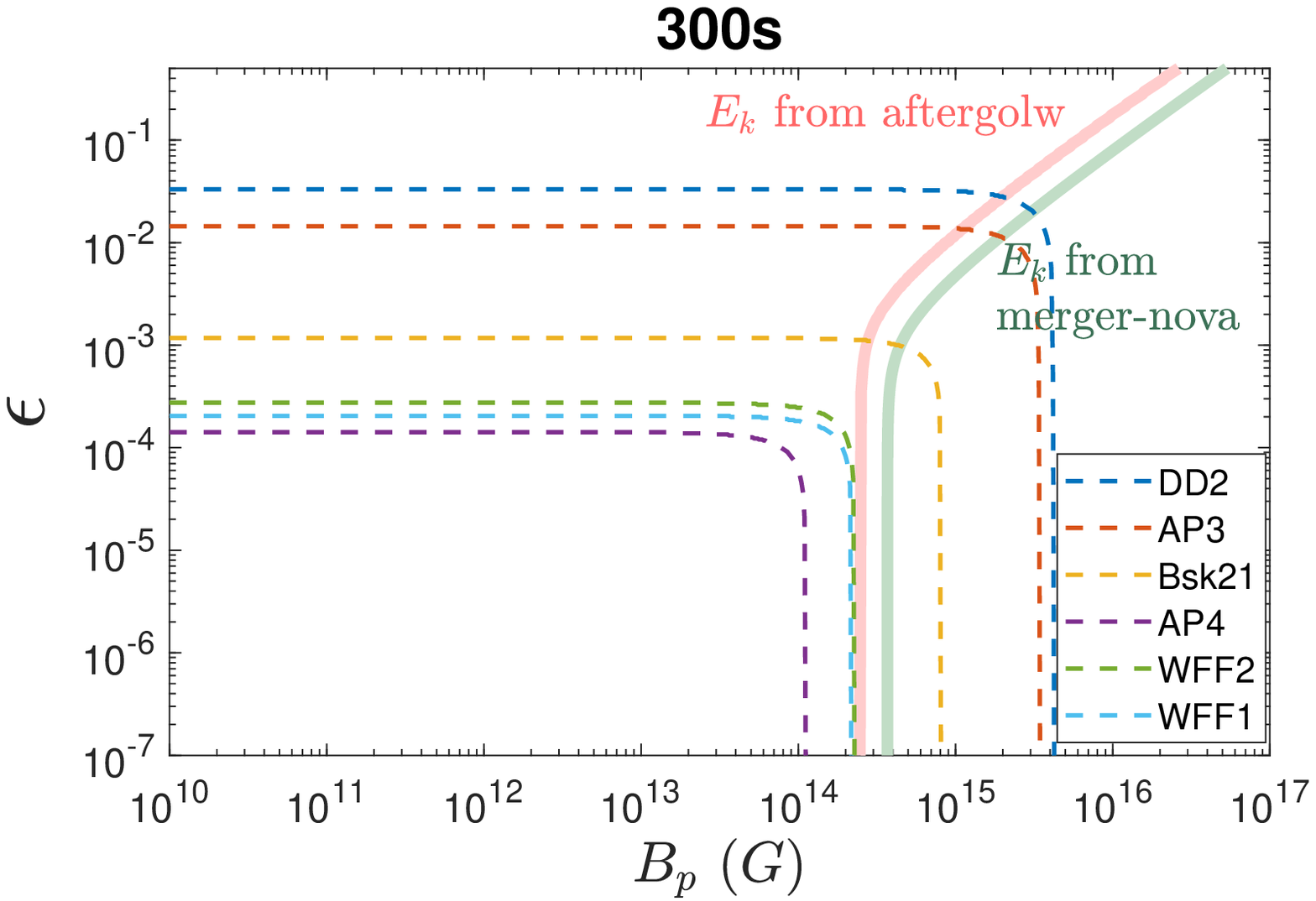}} \\
\resizebox{115mm}{!}{\includegraphics[]{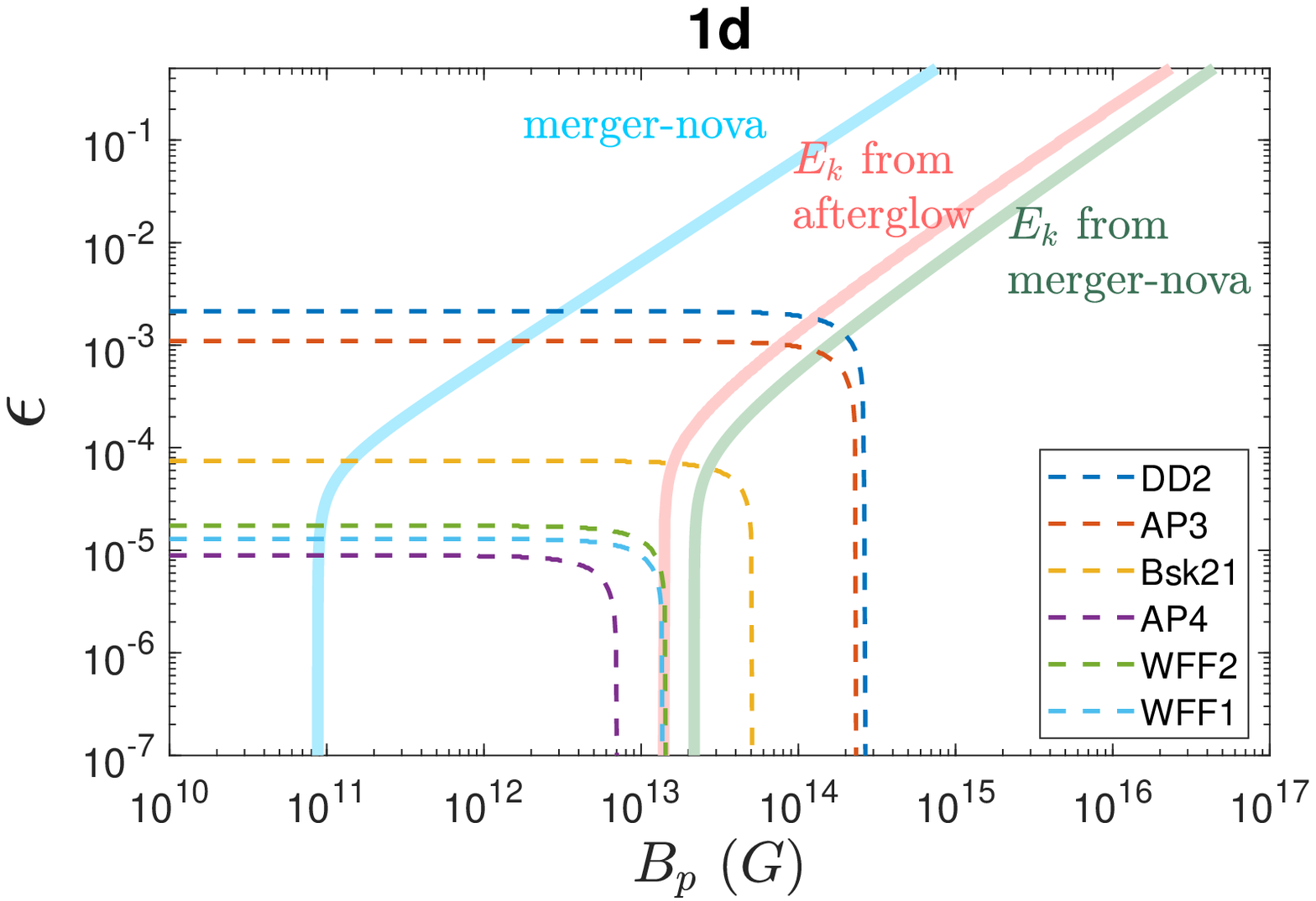}} \\
\resizebox{115mm}{!}{\includegraphics[]{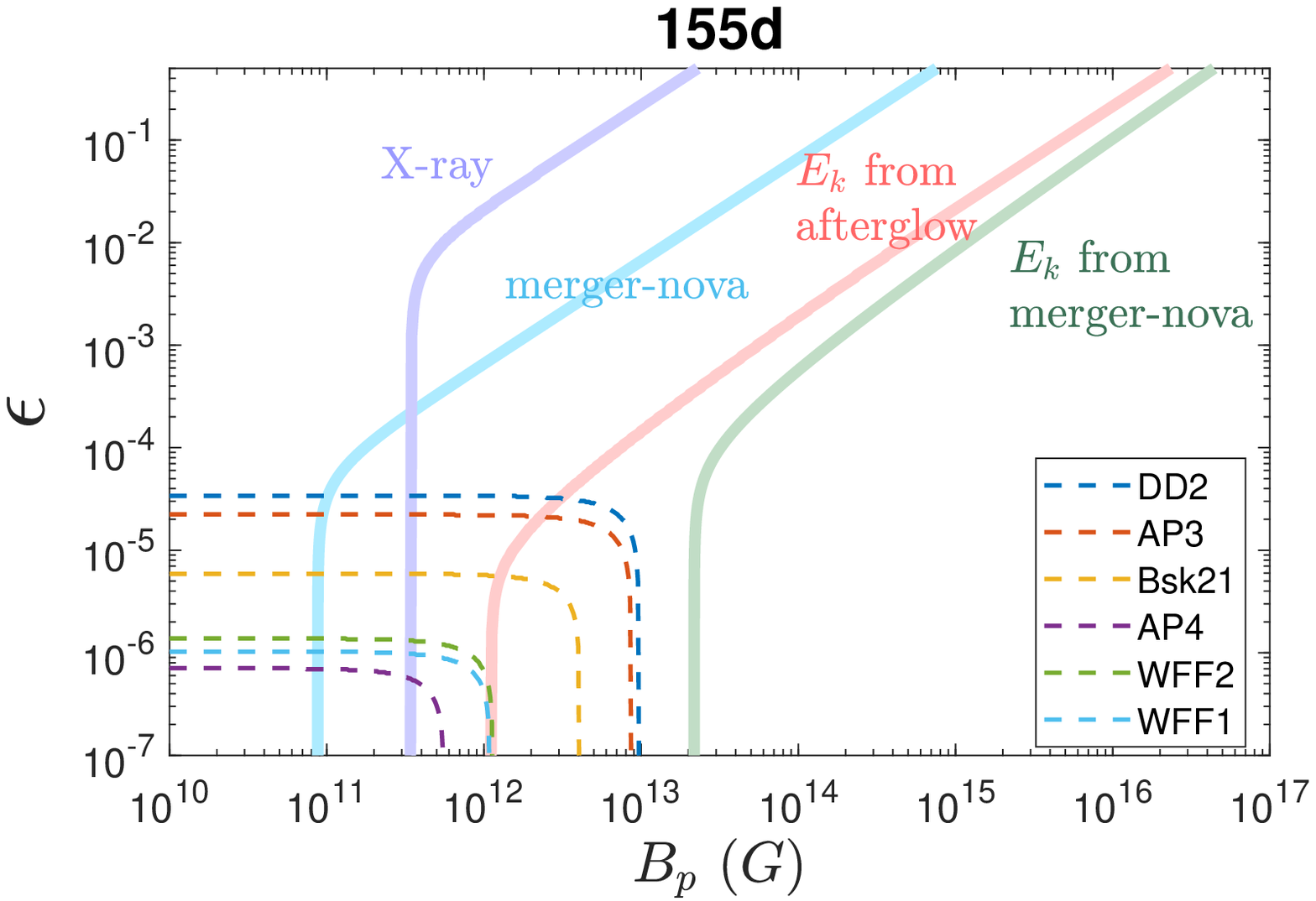}}
\end{tabular}
\caption{Same as Figure 3, but for concrete EoSs studied in this paper.} 
\label{fig:eos}
\end{center}
\end{figure*}

Some of our selected EoSs, e.g. WFF1, WFF2, AP4, BSK21, AP3 and DD2, are supposed to support an SMNS. In principle, with a certain EoS adopted, the properties of the SMNS could be constrained more precisely (see Figure \ref{fig:eos}). If the lifetime of SMNS is 300 s or longer,  without violating the constraints from EM observations, we should have  $B_p<2.1 \times 10^{14} G$ and $\epsilon<2.0\times 10^{-4}$ for WFF1, $B_p<2.2 \times 10^{14}G$ and $\epsilon<2.8\times 10^{-4}$ for WFF2, $B_p<1.1\times 10^{14}G$ and $\epsilon<1.4 \times 10^{-4}$ for AP4, $B_p<7.9 \times 10^{14} G$ and $\epsilon<1.2\times 10^{-3}$ for BSK21, $B_p<3.4 \times 10^{15}G$ and $\epsilon<1.4\times 10^{-2}$ for AP3, $B_p<4.2\times 10^{15}G$ and $\epsilon<3.3 \times 10^{-2}$ for DD2. If the lifetime of SMNS is 1 d or longer,  without violating the EM observational constraints, we should have  $B_p<8.7 \times 10^{10} G$ and $\epsilon<1.3\times 10^{-5}$ for WFF1, $B_p<8.7 \times 10^{10}G$ and $\epsilon<1.7\times 10^{-5}$ for WFF2, $B_p<8.7\times 10^{10}G$ and $\epsilon<8.9 \times 10^{-6}$ for AP4, $B_p<1.5 \times 10^{11} G$ and $\epsilon<7.4\times 10^{-5}$ for BSK21, $B_p<1.6 \times 10^{12}G$ and $\epsilon<1.1\times 10^{-3}$ for AP3 and $B_p<3.2\times 10^{12}G$ and $\epsilon<2.1 \times 10^{-3}$ for DD2. If the lifetime of SMNS is 155 d or longer,  without violating the EM observational constraints, we should have  $B_p<8.5 \times 10^{10} G$ and $\epsilon<1.0\times 10^{-6}$ for WFF1, $B_p<8.5 \times 10^{10}G$ and $\epsilon<1.4\times 10^{-6}$ for WFF2, $B_p<8.5\times 10^{10}G$ and $\epsilon<7.1 \times 10^{-7}$ for AP4, $B_p<8.7 \times 10^{10} G$ and $\epsilon<5.9\times 10^{-6}$ for BSK21, $B_p<9.3 \times 10^{10}G$ and $\epsilon<2.2\times 10^{-5}$ for AP3 and $B_p<1.0\times 10^{11}G$ and $\epsilon<3.4 \times 10^{-5}$ for DD2. The constraints on $B_p$ mainly come from the EM observations, which are roughly consistent with (slightly looser than) the constraints derived in our previous work \citep{ai18}. In this paper, we only used the peak luminosity (or total kinetic energy) rather than the full lightcurve (used in \cite{ai18}) to constrain the parameters. The ellipticity $\epsilon$, which was a free parameter in \cite{ai18}, is now constrained by the lifetime of the SMNS.

\section{Conclusions and Discussion}

A tight constraint on the NS maximum mass $M_{\rm TOV}$ is helpful to constrain the NS EOS. Before GW170817, the constraints on $M_{\rm TOV}$ mainly comes from the observations of Galactic pulsars. Some massive pulsars have been observed, e.g. PSR J1614-2230 with $1.97 \pm 0.04M_{\odot}$ \citep{demorest10} and PSR J0348+0432 with $2.01 \pm 0.04M_{\odot}$ \citep{antoniadis13}), which set a lower limit to $M_{\rm TOV}$ around $\sim 2M_{\odot}$. Recently, the most massive NS PSR J0740+6620 was measured to have a mass $2.14^{+0.10}_{-0.09}~M_{\odot}$ at $68.3\%$ confidence level \citep{cromartie19}. This sets an even more stringent lower limit to $M_{\rm TOV}$.

NS-NS merger events could in principle give tighter constraints on $M_{\rm TOV}$ if the merger product can be unambiguously identified. Unfortunately, without post-merger GW signal (which may not be obtained in the near future), the EM signals are not clean enough to draw definite conclusions. For the case of GW170817, even though the existence of GRB 170817A $\sim 1.7$ s later was regarded by some authors as evidence of the formation of a BH before the onset of the GRB, some other authors argued for a long-lived NS remnant that may exist for an extended period of time. As a result, we cannot place a constraint on $M_{\rm TOV}$. Rather, in this paper, we discuss the range of $M_{\rm TOV}$ for different assumed merger products. We applied two approaches: case studies for 10 individual EoSs and an EoS-independent approach adopting some universal relations. We reached the following self-consistent results: If the merger product was a short-lived HMNS, one has $M_{\rm TOV}<2.09^{+0.11}_{-0.09}(^{+0.06}_{-0.04})M_{\odot}$; If the merger product was a long-lived SMNS, the constraint should be $2.09^{+0.11}_{-0.09}(^{+0.06}_{-0.04})M_{\odot} \leq M_{\rm TOV}<2.43^{+0.10}_{-0.08}(^{+0.06}_{-0.04})M_{\odot}$; If the merger product was a stable NS, the constraints should be $M_{\rm TOV} \geq 2.43^{+0.10}_{-0.08}(^{+0.06}_{-0.04})M_{\odot}$. The quoted uncertainties are at the 2$\sigma$ (1$\sigma$) level.

If the merger remnant is a long-lived MNS, the next question is whether the MNS can survive for a desired period of time, e.g. 300 s, 1 d or 155 d, to interpret various observations. This depends on the spindown history of the remnant, which critically depends on two NS parameters, $B_p$ that defines the dipole spindown and $\epsilon$ that defines the secular GW spindown. For an SNS remnant, this is never a problem. For an SMNS remnant, the survival time actually constrain $B_p$ and $\epsilon$ to be smaller than certain values. We have derived these constraints for the case of GW170817 for different EoSs (which have different $M_{\rm TOV}$ and hence require different $\chi$ values). These constraints, together with those posed from the EM observations, define the parameter space in the $(B_p, \epsilon)$ plane that satisfies the desired lifetime. In general, we find that for any EoS that forms an SMNS in the case of GW170817, without violating the EM observational constraints, there always exist a set of ($B_p, \epsilon$) parameters that makes the SMNS survive for 300 s, 1 d, 155 d or even longer. {In particular, for EoSs in Table 1 with $M_{\rm TOV}$ from $2.14M_{\odot}$ to $2.42M_{\odot}$, we have $\epsilon\lesssim 1.4\times10^{-4}-3.3\times10^{-2}$ and $B_p\lesssim (1.1\times10^{14}-4.2\times10^{15})$G if the SMNS survives for 300 s; $\epsilon\lesssim 8.9\times10^{-6}-2.1\times10^{-3}$ and $B_p\lesssim(8.7\times10^{10}-3.2\times10^{12})$G  if the SMNS survives for 1 d; and $\epsilon\lesssim 7.1\times10^{-7}-3\times1.0\times10^{-6}$ and $B_p\lesssim(8.5\times10^{10}-9.3\times10^{12})$G if the SMNS survives for 155 d.}

Future joint GW/EM observational campaigns of NS-NS merger events may identify more definite observational criteria to identify the nature of the merger remnants. The similar approach proposed in this paper can be applied in those events, which will lead to tighter constraints on $M_{\rm TOV}$ and NS EoS.

\acknowledgments
We thank Luciano Rezzolla and Masaru Shibata for helpful discussion and the anonymous referee for useful comments.
HG acknowleges support by National Natural Science Foundation of China under Grant No. 11722324, 11603003, 11633001 and 11690024, the Strategic Priority Research Program of the Chinese Academy of Sciences, Grant No. XDB23040100 and the Fundamental Research Funds for the Central Universities.

\end{document}